\title{Construction of two SD Codes}
\author{
{\rm Mario Blaum}\\
IBM Almaden Research Center \\
San Jose, CA 95120
\and
{\rm James S. Plank}\\
EECS Department \\
University of Tennessee \\
Knoxville, TN 37996
} 

\documentclass[12pt]{article}
\usepackage{latexsym}
\usepackage{multirow}
\usepackage{color}
 \newtheorem{theo}{Theorem}[section]

 \newtheorem{ex}{Example}[section]

\newtheorem{COROLLARY}{\indent Corollary}

\newtheorem{EXAMPLE}{\indent Example}

\newtheorem{THEOREM}{\indent Theorem}

\newtheorem{REMARK}{\indent Remark}

\newcommand{\fullstop}{\hspace{-0.85em} {\bf .}}

\newcommand{\ue}{\mbox{$\underline{e}$}}

\newcommand{\uc}{\mbox{$\underline{c}$}}

\newcommand{\la}{\mbox{$\leftarrow$}}

\newcommand{\al}{\mbox{$\alpha$}}

\newcommand{\eq}{\mbox{$\, =\,$}}

\newcommand{\qed}{\hfill$\Box$\\[1ex]}
\newcommand{\pf}{{\bf Proof: }}

\newcommand{\xor}{\mbox{$\,\oplus\,$}}

\newcommand{\C}{\mbox{${\cal C}$}}

\newcommand{\cO}{\mbox{${\cal O}$}}

\newcommand{\br}{\\ }
\newcommand{\ce}{\begin{center}}
\newcommand{\cen}{\end{center}}

\newcommand{\ipb}{\begin{description}}
\newcommand{\ipn}{\end{description}}

\newcommand{\qb}{\begin{quote}}
\newcommand{\qn}{\end{quote}}

\newcommand{\tp}{\begin{titlepage}}
\newcommand{\tpn}{\end{titlepage}}
\newcommand{\zb}{\begin{figure}[hbtp]}
\newcommand{\zn}{\end{figure}}

\newcommand{\EQX}[1]{\begin{equation}\label{#1}}
\newcommand{\ENX}{\end{equation}}
\newcommand{\EQL}{\begin{eqnarray*}}
\newcommand{\ENL}{\end{eqnarray*}}
\newcommand{\EQLX}[1]{\begin{eqnarray}\label{#1}}
\newcommand{\ENLX}{\end{eqnarray}}

\newcommand{\open}{\begin{document}}
\newcommand{\close}{\end{document}}
\newcommand{\lfcr}[1]{\br\hspace*{#1em}}
\newenvironment{mat}[1]
{\left[\begin{array}{#1}}{\end{array}\right]}
\newcommand{\GAMMA}{\Gamma}
\newcommand{\DELTA}{\Delta}
\newcommand{\THETA}{\Theta}
\newcommand{\LAMBDA}{\Lambda}
\newcommand{\XI}{\Xi}
\newcommand{\PI}{\Pi}
\newcommand{\SIGMA}{\Sigma}
\newcommand{\UPSILON}{\Upsilon}
\newcommand{\PHI}{\Phi}
\newcommand{\PSI}{\Psi}
\newcommand{\OMEGA}{\Omega}
\newcommand{\bldgreek}[1]{\mbox{\boldmath $#1$}}
\newcommand{\bldbeta}{\bldgreek{\beta}}
\newcommand{\bldgamma}{\bldgreek{\gamma}}
\newcommand{\blddelta}{\bldgreek{\delta}}
\newcommand{\bldepsilon}{\bldgreek{\epsilon}}
\newcommand{\bldvarepsilon}{\bldgreek{\varepsilon}}
\newcommand{\bldzeta}{\bldgreek{\zeta}}
\newcommand{\bldeta}{\bldgreek{\eta}}
\newcommand{\bldtheta}{\bldgreek{\theta}}
\newcommand{\bldvartheta}{\bldgreek{\vartheta}}
\newcommand{\bldiota}{\bldgreek{\iota}}
\newcommand{\bldkappa}{\bldgreek{\kappa}}
\newcommand{\bldlambda}{\bldgreek{\lambda}}
\newcommand{\bldmu}{\bldgreek{\mu}}
\newcommand{\bldnu}{\bldgreek{\nu}}
\newcommand{\bldxi}{\bldgreek{\xi}}
\newcommand{\bldpi}{\bldgreek{\pi}}
\newcommand{\bldvarpi}{\bldgreek{\varpi}}
\newcommand{\bldrho}{\bldgreek{\rho}}
\newcommand{\bldvarrho}{\bldgreek{\varrho}}
\newcommand{\bldsigma}{\bldgreek{\sigma}}
\newcommand{\bldvarsigma}{\bldgreek{\varsigma}}
\newcommand{\bldtau}{\bldgreek{\tau}}
\newcommand{\bldupsilon}{\bldgreek{\upsilon}}
\newcommand{\bldphi}{\bldgreek{\phi}}
\newcommand{\bldvarphi}{\bldgreek{\varphi}}
\newcommand{\bldchi}{\bldgreek{\chi}}
\newcommand{\bldpsi}{\bldgreek{\psi}}
\newcommand{\bldomega}{\bldgreek{\omega}}
\begin{document}
\parindent=10pt
\maketitle
\begin{abstract}
SD codes are erasure codes that address
the mixed failure mode of current RAID systems.  Rather than dedicate entire
disks to erasure coding, as done in RAID-5, RAID-6 and Reed-Solomon coding,
an SD code dedicates entire disks, plus individual sectors to erasure coding.
The code then tolerates combinations of disk and sector errors, rather
than solely disk errors.  It is been an open problem to construct general
codes that have the SD property, and previous work has relied on Monte
Carlo searches.  In this paper, we present two general constructions that
address the cases with one disk and two sectors, and two disks and two sectors.
Additionally, we make an observation about shortening SD codes that allows us to
prune Monte Carlo searches.
\vspace{.3cm}

\end{abstract}

\section{Introduction}
\label{Introduction}

The motivation and description of SD codes is presented in early work
by Plank, Blaum and Hafner~\cite{pbh}.  In this work, we assume that the reader has
read that paper or the follow-on paper~\cite{pb}.

We use the following nomenclature to describe an SD code:

\begin{list}{\mbox{$\bullet$}}{\setlength{\parsep}{-4pt}}
\item $n$: The total number of disks in a disk array.
\item $m$: The total number of disks dedicated to fault-tolerance.
\item $s$: The total number of additional sectors per stripe dedicated to fault-tolerance.
\item $r$: The total number of sectors per disk in a stripe.
\item $GF(2^w)$: The Galois Field which defines the arithmetic.
\item $H$: An $(mr+s) \times (nr)$ parity check matrix.
\end{list}

The parity check matrix has a specific format:

\begin{list}{\mbox{$\bullet$}}{\setlength{\parsep}{-4pt}}
\item For row~$i < mr$, the only non-zero elements are in columns~$\left(\frac{i}{r}\right)n$
through~$\left(\frac{i}{r}\right)(n+1)-1$.  The fractions employ integer division.
\item For row~$mr \le i < mr+s$, all elements are non-zero.
\end{list}

Each block in the stripe has a corresponding column of the parity check matrix.
In particular, block~$i$ of disk~$j$ corresponds to column~$ni+j$.
The code is SD if it tolerates any combination of~$m$ disk failures and~$s$
additional sector failures.

General constructions of SD codes have been heretofore limited.  Blaum,
Hafner and Hetlzer have given constructions when~$s=1$~\cite{bhh},
and Blaum has presented a construction for~$m=1$ and~$s=2$.  In
Section~\ref{construction} of this paper, we present this code again,
but with a simpler proof of the SD property.  In Section~\ref{construction2},
we present a construction for~$m=2$ and~$s=2$.  Finally, in Section~\ref{shorten},
we make an observation on SD codes that allows us to prune searches for
further constructions.

\section{Construction of an SD code with $m=1$ and $s=2$}
\label{construction}

Here we repeat the construction given in~\cite{b}, but we give a
simpler proof.

Consider the field $GF(2^w)$ and let
$\al$ be an element in $GF(2^w)$.
The (multiplicative) order
of $\al$, denoted $\cO(\al)$, is the minimum $\ell$, $0<\ell$, such that
$\al^{\ell}\eq 1$. If $\al$ is a primitive element~\cite{ms}, then
$\cO(\al)\eq 2^w-1$. To each element $\al\in GF(2^w)$, there is an
associated (irreducible) minimal polynomial~\cite{ms} that we denote $f_{\al}(x)$.

Let $\al\in GF(2^w)$ and $rn\leq \cO(\al)$. Consider the $(r+2)\times rn$ parity-check matrix

\begin{eqnarray}
\label{pcSD}
\left(
\begin{array}{cccc|cccc|c|cccc}
\uc_0&\uc_1&\ldots&\uc_{n-1}&\uc_{n}&\uc_{n+1}&\ldots&\uc_{2n-1}&\ldots
&\uc_{(r-1)n}&\uc_{(r-1)n+1}&\ldots&\uc_{rn-1}\\
\end{array}
\right)
\end{eqnarray}
where $\uc_i$ denotes a column of length $r+2$, and, if $\ue_i$
denotes an $r\times 1$ vector whose coordinates are zero except for
coordinate $i$, which is 1, then, for $0\leq i\leq r-1$,

\begin{eqnarray}
\label{pcSDc}
\uc_{in},\uc_{in+1},\ldots ,\uc_{(i+1)n-1}&=&\left(
\begin{array}{cccccc}
\ue_i &\ue_i &\ldots &\ue_i & \ldots &\ue_i \\
\al^{in}&\al^{in+1}&\ldots &\al^{in+j}&\ldots &\al^{(i+1)n-1}\\
\al^{2in}&\al^{2in-1}&\ldots &\al^{2in-j}&\ldots &\al^{(2i-1)n+1}\\
\end{array}
\right)
\end{eqnarray}

We denote  as $\C(r,n,1;f_{\al}(x))$ the $[rn,r(n-1)-2]$ code over
$GF(2^w)$ whose parity-check  matrix is given
by~(\ref{pcSD}) and~(\ref{pcSDc}).

\begin{ex}
\label{excode}
{\em
Consider the finite field $GF(16)$ and let $\al$ be a primitive
element, i.e., $\cO(\al)\eq 15$. Then, the parity-check matrix of
$\C(3,5,1;f_{\al}(x))$ is given by

$$
\left(
\begin{array}{ccccc|ccccc|ccccc}
1&1&1&1&1&0&0&0&0&0&0&0&0&0&0\\
0&0&0&0&0&1&1&1&1&1&0&0&0&0&0\\
0&0&0&0&0&0&0&0&0&0&1&1&1&1&1\\
1&\al&\al^2&\al^3&\al^4&\al^5&\al^6&\al^7&\al^8&\al^9&\al^{10}&\al^{11}&\al^{12}&\al^{13}&\al^{14}\\
1&\al^{14}&\al^{13}&\al^{12}&\al^{11}&
\al^{10}&\al^9&\al^8&\al^7&\al^6&\al^{5}&\al^{4}&\al^{3}&\al^{2}&\al\\
\end{array}
\right)
$$

Similarly, the parity-check matrix of
$\C(5,3,1;f_{\al}(x))$ is given by

$$
\left(
\begin{array}{ccc|ccc|ccc|ccc|ccc}
1&1&1&0&0&0&0&0&0&0&0&0&0&0&0\\
0&0&0&1&1&1&0&0&0&0&0&0&0&0&0\\
0&0&0&0&0&0&1&1&1&0&0&0&0&0&0\\
0&0&0&0&0&0&0&0&0&1&1&1&0&0&0\\
0&0&0&0&0&0&0&0&0&0&0&0&1&1&1\\
1&\al&\al^2&\al^3&\al^4&\al^5&\al^6&\al^7&\al^8&\al^9&\al^{10}&\al^{11}&\al^{12}&\al^{13}&\al^{14}\\
1&\al^{14}&\al^{13}&\al^{6}&\al^{5}&
\al^{4}&\al^{12}&\al^{11}&\al^{10}&\al^3&\al^{2}&\al &\al^{9}&\al^{8}&\al^7\\
\end{array}
\right)
$$
}
\end{ex}

Let us point out that the construction of this type of codes is valid
also over the ring of polynomials modulo $M_p(x)\eq 1+x+\cdots
+x^{p-1}$, $p$ a prime number, as done with the Blaum-Roth (BR)
codes~\cite{br}. In that case, $\cO(\al)\eq p$, where $\al^{p-1}\eq
1+\al+\cdots +\al^{p-2}$. The construction
proceeds similarly, and we denote it $\C(r,n,1;M_p(x))$. Utilizing
the ring modulo $M_p(x)$ allows for XOR operations at the encoding
and the decoding without look-up tables in a finite field, which is
advantageous in erasure decoding~\cite{br}. It is well known that
$M_p(x)$ is irreducible if and only if 2 is primitive in
$GF(p)$~\cite{ms}.

\begin{ex}
\label{exM17}
{\em
Consider the ring of polynomials modulo $M_{17}(x)$ and let $\al$ be an
element in the ring such that $\al^{16}\eq
1+\al+\cdots +\al^{15}$, thus, $\cO(\al)\eq 17$ (notice, $M_{17}(x)$
is reducible). Then, the parity-check matrix of
$\C(4,4,1;M_{17}(x))$ is given by

$$
\left(
\begin{array}{cccc|cccc|cccc|cccc}
1&1&1&1&0&0&0&0&0&0&0&0&0&0&0&0\\
0&0&0&0&1&1&1&1&0&0&0&0&0&0&0&0\\
0&0&0&0&0&0&0&0&1&1&1&1&0&0&0&0\\
0&0&0&0&0&0&0&0&0&0&0&0&1&1&1&1\\
1&\al&\al^2&\al^3&\al^4&\al^5&\al^6&\al^7&\al^8&\al^9&\al^{10}&\al^{11}&\al^{12}&\al^{13}&\al^{14}&\al^{15}\\
1&\al^{16}&\al^{15}&\al^{14}&\al^{8}&\al^{7}&\al^6&\al^5&\al^{16}&\al^{15}&\al^{14}&\al^{13}&\al^{7}&\al^{6}&\al^5&\al^4\\
\end{array}
\right)
$$
}
\end{ex}

We have the following theorem:

\begin{theo}
\label{theoSD}
{\em
Codes $\C(r,n,1;f_{\al}(x))$ and $\C(r,n,1;M_p(x))$ are SD codes.
}
\end{theo}

\pf We break the proof into two cases.  In the first, the two sector errors
occur on the same row of the stripe.  In this case, we focus solely
on three columns of the parity check matrix that share non-zero entries
in one of the first~$r$ rows.  Put another way,
this will happen if and only if, for any $0\leq i\leq r-1$ and $0\leq j_0<j_1<j_2\leq n-1$,

\begin{eqnarray*}
\det\left(
\begin{array}{ccc}
1&1&1\\
\al^{in+j_0}&\al^{in+j_1}&\al^{in+j_2}\\
\al^{2in-j_0}&\al^{2in-j_1}&\al^{2in-j_2}\\
\end{array}
\right)&\neq &0
\end{eqnarray*}

But the determinant of this $3\times 3$ matrix can be easily transformed into a Vandermonde
determinant on $\al^{j_0}$, $\al^{j_1}$ and $\al^{j_2}$ times a power of
$\al$, so it is invertible in a field and also in the ring of
polynomials modulo $M_p(x)$~\cite{br}.

In the second case, the two sector failures occur in different rows of
the stripe. In this case, we must prove
to prove that if we have two erasures in locations $i$
and $j$ of row $\ell$, and two erasures
in locations $i$ and $j'$ of row $\ell'$, $0\leq i,j,j'\leq n-1$,
$j,j'\neq i$,
$0\leq \ell<\ell'\leq r-1$, then

\begin{eqnarray*}
\det\left(
\begin{array}{cccc}
1&1&0&0\\
0&0&1&1\\
\al^{\ell n+i}&\al^{\ell n+j}&\al^{\ell' n+i}&\al^{\ell' n+j'}\\
\al^{2\ell n-i}&\al^{2\ell n-j}&\al^{2\ell' n-i}&\al^{2\ell' n-j'}\\
\end{array}
\right)&\neq &0
\end{eqnarray*}

After some row manipulation, the inequality above holds if and only
if

\begin{eqnarray*}
\det\left(
\begin{array}{cc}
\al^{\ell n+i}\left(1\xor\al^{j-i}\right)&\al^{\ell' n+i}\left(1\xor\al^{j'-i}\right)\\
\al^{2\ell n-j}\left(1\xor\al^{j-i}\right)&\al^{2\ell' n-j'}\left(1\xor\al^{j'-i}\right)\\
\end{array}
\right)&\neq &0.
\end{eqnarray*}
Both $1\xor\al^{j-i}$ and $1\xor\al^{j'-i}$ are invertible in $GF(2^w)$
since $1\leq j-i,j'-i<\cO(\al)$,
and the same is true in the
polynomials modulo $M_p(x)$~\cite{br}, thus, the inequality above is
satisfied if and only if

\begin{eqnarray*}
\det\left(
\begin{array}{cc}
\al^{-(\ell'-\ell)n}&1\\
1&\al^{2(\ell'-\ell)n+j-j'}\\
\end{array}
\right)&\eq &1\xor \al^{(\ell'-\ell)n+j-j'}.
\end{eqnarray*}

Redefining $\ell\,\la\,\ell'-\ell$ and $j\,\la\,j'-j$, we have
$1\leq\ell\leq r-1$ and $-(n-1)\leq j\leq n+1$.
Thus, $\al^{\ell n+j}\neq 1$, since
\begin{eqnarray*}
1\leq \ell n+j\leq (r-1)n+n-1\eq rn-1\leq \cO(\al)-1.
\end{eqnarray*}
\qed

By Theorem~\ref{theoSD}, the codes $\C(3,5,1;f_{\al}(x))$ and $\C(5,3,1;f_{\al}(x))$ in
Example~\ref{excode} are SD, as well as code $\C(4,4,1;M_{17}(x))$
in Example~\ref{exM17}.

\section{Construction of an SD code with $m=2$ and $s=2$}
\label{construction2}
Let $\al\in GF(2^w)$ and $rn\leq \cO(\al)$. Consider the $(2r+2)\times rn$ parity-check matrix

\begin{eqnarray}
\label{pcSD2}
\left(
\begin{array}{cccc|cccc|c|cccc}
\uc_0&\uc_1&\ldots&\uc_{n-1}&\uc_{n}&\uc_{n+1}&\ldots&\uc_{2n-1}&\ldots
&\uc_{(r-1)n}&\uc_{(r-1)n+1}&\ldots&\uc_{rn-1}\\
\end{array}
\right)
\end{eqnarray}
where $\uc_{in+j}$ denotes a column of length $2r+2$, and, if $\ue_{in+j}$
denotes a $2r\times 1$ vector whose coordinates are zero except for
coordinates $2i$ and $2i+1$, which are 1 and $\al^{j}$
respectively, then, for $0\leq i\leq r-1$,

\begin{eqnarray}
\label{pcSDc2}
\uc_{in},\uc_{in+1},\ldots ,\uc_{(i+1)n-1}&=&\left(
\begin{array}{cccccc}
\ue_{in} &\ue_{in+1} &\ldots &\ue_{in+j} & \ldots &\ue_{(i+1)n-1} \\
\al^{3in}&\al^{3in-1}&\ldots &\al^{3in-j}&\ldots &\al^{3in-(n-1)}\\
\al^{2in}&\al^{2(in+1)}&\ldots &\al^{2(in+j)}&\ldots &\al^{2((i+1)n-1)}\\
\end{array}
\right)
\end{eqnarray}

We denote  as $\C(r,n,2;f_{\al}(x))$ the $[rn,r(n-2)-2]$ code over
$GF(q)$ whose parity-check  matrix is given
by~(\ref{pcSD2}) and~(\ref{pcSDc2}).

Let us illustrate the construction of $\C(r,n,2;f_{\al}(x))$ with an example.

\begin{ex}
\label{excode3}
{\em
As in Example~\ref{excode}, consider the finite
field $GF(16)$ and let $\al$ be a primitive
element, i.e., $\cO(\al)\eq 15$. Then, the parity-check matrix of
$\C(3,5,2;f_{\al}(x))$ is given by

$$
\left(
\begin{array}{ccccc|ccccc|ccccc}
1&1&1&1&1&0&0&0&0&0&0&0&0&0&0\\
1&\al&\al^2&\al^3&\al^4&0&0&0&0&0&0&0&0&0&0\\
0&0&0&0&0&1&1&1&1&1&0&0&0&0&0\\
0&0&0&0&0&1&\al&\al^2&\al^3&\al^4&0&0&0&0&0\\
0&0&0&0&0&0&0&0&0&0&1&1&1&1&1\\
0&0&0&0&0&0&0&0&0&0&1&\al&\al^2&\al^3&\al^4\\
1&\al^{14}&\al^{13}&\al^{12}&\al^{11}&
1&\al^{14}&\al^{13}&\al^{12}&\al^{11}&
1&\al^{14}&\al^{13}&\al^{12}&\al^{11}\\
1&\al^2&\al^4&\al^6&\al^8&\al^{10}&\al^{12}&\al^{14}&\al&\al^3&\al^{5}&\al^{7}&\al^{9}&\al^{11}&\al^{13}\\
\end{array}
\right)
$$

Similarly, the parity-check matrix of
$\C(5,3,2;f_{\al}(x))$ is given by

$$
\left(
\begin{array}{ccc|ccc|ccc|ccc|ccc}
1&1&1&0&0&0&0&0&0&0&0&0&0&0&0\\
1&\al&\al^2&0&0&0&0&0&0&0&0&0&0&0&0\\
0&0&0&1&1&1&0&0&0&0&0&0&0&0&0\\
0&0&0&1&\al&\al^2&0&0&0&0&0&0&0&0&0\\
0&0&0&0&0&0&1&1&1&0&0&0&0&0&0\\
0&0&0&0&0&0&1&\al&\al^2&0&0&0&0&0&0\\
0&0&0&0&0&0&0&0&0&1&1&1&0&0&0\\
0&0&0&0&0&0&0&0&0&1&\al&\al^2&0&0&0\\
0&0&0&0&0&0&0&0&0&0&0&0&1&1&1\\
0&0&0&0&0&0&0&0&0&0&0&0&1&\al&\al^2\\
1&\al^{14}&\al^{13}&\al^{9}&\al^{8}&
\al^{7}&\al^3&\al^2&\al&\al^{12}&\al^{11}&\al^{10}&\al^{6}&\al^{5}&\al^4\\
1&\al^2&\al^4&\al^6&\al^8&\al^{10}&\al^{12}&\al^{14}&\al&\al^3&\al^{5}&\al^{7}&\al^{9}&\al^{11}&\al^{13}\\
\end{array}
\right)
$$
}
\end{ex}

Next we prove the following theorem:

\begin{theo}
\label{theoSD2}
{\em
Codes $\C(r,n,2;f_{\al}(x))$ and
$\C(r,n,2;M_p(x))$ are SD codes.
}
\end{theo}

\pf Notice
that 4 erasures in the same row will always be corrected. In effect,
based on the
parity-check matrix of the code, this will happen if and only if, for
any $0\leq i\leq r-1$ and $0\leq t_0<t_1<t_2<t_3\leq n-1$,

\begin{eqnarray*}
\det\left(
\begin{array}{cccc}
1&1&1&1\\
\al^{t_0}&\al^{t_1}&\al^{t_2}&\al^{t_3}\\
\al^{3in-t_0}&\al^{3in-t_1}&\al^{3in-t_2}&\al^{3in-t_3}\\
\al^{2(in+t_0)}&\al^{2(in+t_1)}&\al^{2(in+t_2)}&\al^{2(in+t_3)}\\
\end{array}
\right)&\neq &0
\end{eqnarray*}

The determinant of this $4\times 4$ matrix can be easily transformed into a Vandermonde
determinant on $\al^{t_0}$, $\al^{t_1}$, $\al^{t_2}$ and $\al^{t_3}$ times a power of
$\al$, so it is invertible in a field and also in the ring of
polynomials modulo $M_p(x)$~\cite{br}.

Thus, the code will be SD if and only if, given three erasures in locations
$i$, $j$ and $t$ of row $\ell$, and three erasures
in locations $i$, $j$ and $t'$ of row $\ell'$, $0\leq i<j\leq n-1$,
$0\leq t,t'\leq n-1$, $t,t'\neq i$ and $t,t'\neq j$,
$0\leq \ell<\ell'\leq r-1$, then

\begin{eqnarray*}
\det\left(
\begin{array}{cccccc}
1&1&1&0&0&0\\
\al^{t}&\al^{i}&\al^{j}&0&0&0\\
0&0&0&1&1&1\\
0&0&0&\al^{t'}&\al^{i}&\al^{j}\\
\al^{3\ell n-t}&\al^{3\ell n-i}&\al^{3\ell n-j}&\al^{3\ell' n-t'}&\al^{3\ell' n-i}&\al^{3\ell' n-j}\\
\al^{2(\ell n+t)}&\al^{2(\ell n+i)}&\al^{2(\ell n+j)}&\al^{2(\ell' n+t')}&\al^{2(\ell' n+i)}&\al^{2(\ell' n+j)}\\
\end{array}
\right)&\neq &0.
\end{eqnarray*}

After some row manipulation, the inequality above holds if and only
if

\begin{eqnarray*}
\det\left(
\begin{array}{cccc}
\al^{t}(1\xor \al^{i-t})&\al^{t}(1\xor \al^{j-t})&0&0\\
0&0&\al^{t'}(1\xor \al^{i-t'})&\al^{t'}(1\xor \al^{j-t'})\\
\al^{3\ell n-i}(1\xor \al^{i-t})&\al^{3\ell n-j}(1\xor
\al^{j-t})&\al^{3\ell' n-i}(1\xor \al^{i-t'})&\al^{3\ell' n-j}(1\xor \al^{j-t'})\\
\al^{2(\ell n+t)}(1\xor \al^{i-t})^2&\al^{2(\ell n+t)}(1\xor
\al^{j-t})^2&\al^{2(\ell' n+t')}(1\xor \al^{i-t'})^2&\al^{2(\ell' n+t')}(1\xor \al^{j-t'})^2\\
\end{array}
\right)&\neq &0
\end{eqnarray*}
if and only if, taking some common factors in columns and in the
first two rows,

\begin{eqnarray*}
\det\left(
\begin{array}{cccc}
1&1&0&0\\
0&0&1&1\\
\al^{3\ell n-i}&\al^{3\ell n-j}&\al^{3\ell' n-i}&\al^{3\ell' n-j}\\
\al^{2(\ell n+t)}(1\xor \al^{i-t})&\al^{2(\ell n+t)}(1\xor
\al^{j-t})&\al^{2(\ell' n+t')}(1\xor \al^{i-t'})&\al^{2(\ell' n+t')}(1\xor \al^{j-t'})\\
\end{array}
\right)&\neq &0.
\end{eqnarray*}

Some more row manipulation gives that this determinant is nonzero if
and only if

\begin{eqnarray*}
\det\left(
\begin{array}{cc}
\al^{3\ell n-j}(1\xor \al^{j-i})&\al^{3\ell' n-j}(1\xor \al^{j-i})\\
\al^{2(\ell n+t)+i-t}(1\xor \al^{j-i})&\al^{2(\ell' n+t')+i-t'}(1\xor \al^{j-i})\\
\end{array}
\right)&\neq &0
\end{eqnarray*}
if and only if

\begin{eqnarray*}
\det\left(
\begin{array}{cc}
1&\al^{3(\ell'-\ell)}\\
\al^{-2(\ell' -\ell)n+t-t'}&1\\
\end{array}
\right)&\eq &1\xor\al^{(\ell' -\ell)n+t-t'}\neq 0 .
\end{eqnarray*}

Redefining $\ell\,\la\,\ell'-\ell$ and $t\,\la\,t-t'$, we have $1\leq
\ell\leq r-1$ and $-(n-1)\leq t\leq n-1$. Then, $1\xor \al^{\ell
n+t}\neq 0$ reasoning as in Theorem~\ref{theoSD}.
\qed

For instance, consider the finite field $GF(256)$ and
let $\al$ be a primitive
element, i.e., $\cO(\al)\eq 255$. Then, by Theorem~\ref{theoSD2}, code
$\C(51,5,2;f_{\al}(x))$ is SD.

\section{Shortening the Codes in order to Prune Searches}
\label{shorten}

Observe the following: if we have an
SD code consisting of $r\times n$ arrays,
we consider the subcode of $r\times n$ arrays such that the last
$r-r'$ rows are zero, where $r'<r$. These arrays correspond to a
shortening of the original code and are written simply as
$r'\times n$ arrays. Since shortening preserves the
error-correcting properties of the code, the shortened code
consisting of $r'\times n$ arrays is also SD.
This observation allows us to prune the
search as follows:
in~\cite{pb}, we performed Monte Carlo searches to discover SD
codes in cases where we didn't have constructions.  Our methodology is to
construct codes using random coefficients to generate a parity check
matrix for given values of~$n$, $m$, $s$ and $r$ in~$GF(2^w)$.  We
then test to see if the code is SD.  If it is, then we generate a
parity check matrix for~$n$, $m$, $s$ and $r+1$, and test it to see if
the code is SD.  We continue until either~$r$
reaches a threshold, or until the code is not SD.  Because of the
observation above regarding shortening,
when we discover a code that is not SD, we are guaranteed that codes for higher
values of~$r$ are not SD, and therefore we do not have to generate and test them.

\subsection{Acknowledgements}

This work is supported by the National Science Foundation, under
grant CSR-1016636, and by an IBM Faculty Award.

\end{document}